 \documentclass[twocolumn,prl,aps,psfig,showpacs,preprintnumbers,superscriptaddress]{revtex4}
\usepackage{graphicx}
\usepackage{epstopdf}
\usepackage{float}
\usepackage{color}
\usepackage{amsmath}

\begin{document}

\title{NMR Study of the New Magnetic Superconductor CaK(Fe$_{0.951}$Ni$_{0.049}$)$_4$As$_4$: Microscopic Coexistence of Hedgehog Spin-vortex Crystal and Superconductivity}
\author{Q.-P. Ding}
\affiliation{Ames Laboratory, U.S. DOE, and Department of Physics and Astronomy, Iowa State University, Ames, Iowa 50011, USA}
\author{W. R. Meier}
\affiliation{Ames Laboratory, U.S. DOE, and Department of Physics and Astronomy, Iowa State University, Ames, Iowa 50011, USA}
\author{A. E. B\"ohmer}
\affiliation{Ames Laboratory, U.S. DOE, and Department of Physics and Astronomy, Iowa State University, Ames, Iowa 50011, USA}
\author{S. L. Bud'ko}
\affiliation{Ames Laboratory, U.S. DOE, and Department of Physics and Astronomy, Iowa State University, Ames, Iowa 50011, USA}
\author{P. C. Canfield}
\affiliation{Ames Laboratory, U.S. DOE, and Department of Physics and Astronomy, Iowa State University, Ames, Iowa 50011, USA}
\author{Y. Furukawa}
\affiliation{Ames Laboratory, U.S. DOE, and Department of Physics and Astronomy, Iowa State University, Ames, Iowa 50011, USA}

\date{\today}

\begin{abstract} 

     Coexistence of a new-type antiferromagnetic (AFM) state, the so-called hedgehog spin-vortex crystal (SVC), and superconductivity (SC)  is evidenced by $^{75}$As nuclear magnetic resonance study on single-crystalline CaK(Fe$_{0.951}$Ni$_{0.049}$)$_4$As$_4$.
     The  hedgehog SVC order is clearly demonstrated by the direct observation of the internal magnetic induction along the $c$ axis at the As1 site (close to K) and a zero net internal magnetic induction at the As2 site (close to Ca) below an AFM ordering temperature $T_{\rm N}$ $\sim$ 52 K.
     The nuclear spin-lattice relaxation rate 1/$T_1$ shows a distinct decrease below $T_{\rm c}$ $\sim$ 10 K, providing also unambiguous evidence for the microscopic coexistence.
     Furthermore, based on the analysis of the 1/$T_1$ data, the hedgehog SVC-type spin correlations are found to be enhanced below $T$ $\sim$ 150 K in the paramagnetic state.
      These results indicate the hedgehog SVC-type spin correlations play an important role for the appearance of SC in the new magnetic superconductor.

\end{abstract}

\maketitle


   The relationship between antiferromagnetism (AFM) and superconductivity (SC) has received wide interest in the study of high-temperature SC in  iron-based superconductors.
    Among the iron-based superconducting compounds, those of the 122-type family $Ae$Fe$_2$As$_2$ ($Ae$ = Ca, Ba, Sr, Eu) with a ThCr$_2$Si$_2$-type structure at room temperature attracted most attention \cite{Johnston2010,Canfield2010,Stewart2011}.
     In these systems, by lowering temperature, the crystal structure changes from high-temperature tetragonal ($C_4$ symmetry)  to low-temperature orthorhombic ($C_2$ symmetry) at, or just above, a system-dependent N\'eel temperature $T_{\rm N}$, below which long-range stripe-type AFM order emerges.
     SC in these compounds emerges upon suppression of both the structural and magnetic transitions by application of carrier doping and/or pressure.
    The stripe-type AFM order with the $C_2$ symmetry coexists with the SC phase in various doped $``$122$"$ compounds such as  Ba$_{1-x}$K$_x$Fe$_2$As$_2$,  BaFe$_2$(As$_{1-x}$P$_x$)$_2$, Ba(Fe$_{1-x}$Co$_x$)$_2$As$_2$, and Ba(Fe$_{1-x}$Ru$_x$)$_2$As$_2$ \cite{Urbano2010,Avci2011,Wiesenmayer2011,Iye2012,Li2012,Pratt2009,Ni2010,Ma2012,Laplace2009}, but not in Ca(Fe$_{1-x}$Co$_x$)$_2$As$_2$ \cite{Cui2015}.

     Recently, new magnetic states with the $C_4$-symmetry have attracted much attention \cite{Hoyer2016}.  
      These states are characterized by the wave vectors ${\bf Q}_1$ = ($\pi$,0) and ${\bf Q}_2$  = (0,$\pi$) as in the stripe-ordered state, and can be understood as the superposition of two spin density waves (SDW) ${\bf S(r)}$ = ${\bf M}_1e^{i{\bf Q}_1\cdot{\bf r}}$ +  ${\bf M}_2e^{i{\bf Q}_2\cdot{\bf r}}$ \cite{Hoyer2016,Fernandes2016, O'Halloran2017}. 
    Here, ${\bf M}_1$ and ${\bf M}_2$ are the magnetic order parameters associated with the two wave vectors ${\bf Q}_1$ and ${\bf Q}_2$, respectively. 
      When ${\bf M}_1$ and ${\bf M}_2$ are either parallel or antiparallel, a nonuniform magnetization is produced where the average moment at one lattice site vanishes and staggered-like order appears at the other lattice sites \cite{Fernandes2016}. 
      This so-called charge-spin density wave (CSDW) has been demonstrated to be realized in Sr$_{1-x}$Na$_x$Fe$_2$As$_2$ \cite{Allred2016}, and likely occurs in Ba(Fe$_{1-x}$Mn$_x$)$_2$As$_2$, Ba$_{1-x}$Na$_x$Fe$_2$As$_2$, and Ba$_{1-x}$K$_x$Fe$_2$As$_2$ as well  \cite{Kim2010,Avci2014,Wang2016,Hassinger2012,Bohmer2015,Allred2015,Hassinger2016}.
      A possible coexistence of CSDW and SC is reported in Ba$_{1-x}$Na$_x$Fe$_2$As$_2$ \cite{Avci2014}, and  Ba$_{1-x}$K$_x$Fe$_2$As$_2$ \cite{Wang2016,Bohmer2015,Hassinger2016}.

\begin{figure}[tb]
\includegraphics[width=8cm]{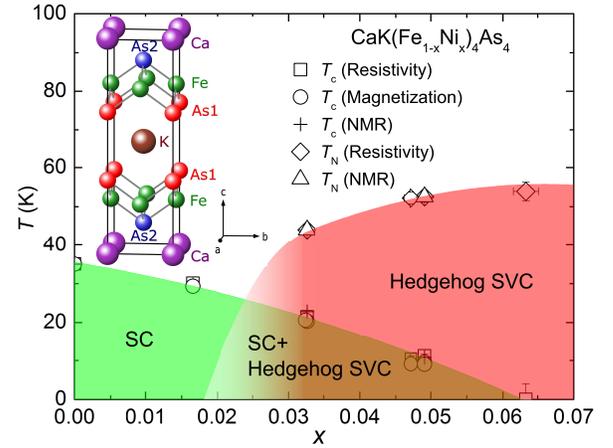} 
\caption{Phase diagram of CaK(Fe$_{1-x}$Ni$_{x}$)$_4$As$_4$. $T_{\rm N}$ and  $T_{\rm c}$ determined from resistivity and magnetization are  from Ref. \onlinecite{Meier20172}. The inset shows the crystal structure of CaKFe$_4$As$_4$ where the two crystallographically inequivalent As sites exist: As1 and As2 sites close to the K and Ca layers, respectively.}
\label{fig:Phasediagram}
\end{figure}

     Very recently, a new magnetic state called $``$hedgehog" spin vortex crystal (SVC) with the $C_4$ symmetry has been identified in the electron doped 1144-type iron pnictide SC CaK(Fe$_{1-x}M_{x}$)$_4$As$_4$ ($M$ = Co or Ni) \cite{Meier20172} (Fig.\ \ref{fig:Phasediagram}).
    The hedgehog SVC state is another double-${\bf Q}$ SDW state in the iron-based systems, in which ${\bf M}_1$ and ${\bf M}_2$ are orthogonal.
Importantly, CaK(Fe$_{1-x}M_{x}$)$_4$As$_4$ ($M$ = Co or Ni) crystallizes through alternate stacking of the Ca and K layers across the Fe$_2$As$_2$ layer as a result of the large ionic radius difference \cite {Meier20172,Iyo2016,Meier2016}.
        The ordering of the Ca and K layers changes the space group from $I$4$/mmm$ in $A$Fe$_2$As$_2$ to $P$4$/mmm$ in CaKFe$_4$As$_4$ (CaK1144).
        Consequently, as shown in the inset of  Fig.\ \ref{fig:Phasediagram}, there are two inequivalent As sites: As1 and As2 sites close to the K and Ca layers, respectively.
          The multiband nature and  two nodeless isotropic superconducting gaps have been revealed in the parent compound by various techniques \cite{ARPES2016,Lochner2017,uSR2017,Cho2017,Jean2017,Iida2017}.
    Whereas an overlap between the hedgehog SVC and the SC regions appears in the phase diagram compiled in Ref. \onlinecite{Meier20172} and shown in Fig.\ \ref{fig:Phasediagram}, it remains an important question whether or not this new AFM state coexists with SC at a microscopic scale.  
    Furthermore, the spin fluctuations in the materials with hedgehog SVC order have never been investigated.
    Here we have carried out  $^{75}$As NMR study on CaK(Fe$_{0.951}$Ni$_{0.049}$)$_4$As$_4$ (4.9$\%$Ni-CaK1144) single crystals ($T_{\rm N}$ $\sim$  52 K, $T_{\rm c}$ $\sim$  10.5 K)  in order to investigate the magnetic and electronic properties from a microscopic point of view.
    The well defined NMR signals from the As1 and As2 sites allow us to determine the temperature dependence of hyperfine fields and magnetic fluctuations at each site separately, providing clear evidence of the coexistence of the hedgehog SVC and SC.


   Single crystals  of 4.9$\%$Ni-CaK1144 for the NMR measurements were grown out of a high-temperature solution rich in transition-metals and arsenic \cite {Meier20172,Meier2016,Meier2017}. 
   NMR measurements of $^{75}$As ($I$ = $\frac{3}{2}$, $\frac{\gamma_{\rm N}}{2\pi}$ = 7.2919 MHz/T, $Q=$ 0.29 barns) nuclei were conducted using a lab-built phase-coherent spin-echo pulse spectrometer. 
   In situ ac magnetic susceptibility ($\chi_{\rm ac}$)  was measured by monitoring the resonance frequency $f$ of the NMR coil tank circuit  as a function of temperature ($T$)  using a network analyzer.
   The $^{75}$As-NMR spectra were obtained by sweeping the magnetic field $H$ at a fixed frequency $f$ = 43.2 MHz.
   The $^{75}$As nuclear spin-lattice relaxation rate 1/$T_{\rm 1}$ was measured with a saturation recovery method \cite{T1}.
   Preliminary results of $^{75}$As NMR spectrum measurements have been reported in the previous paper \cite{Meier20172}.

\begin{figure}[tb]
\includegraphics[width=8.8cm]{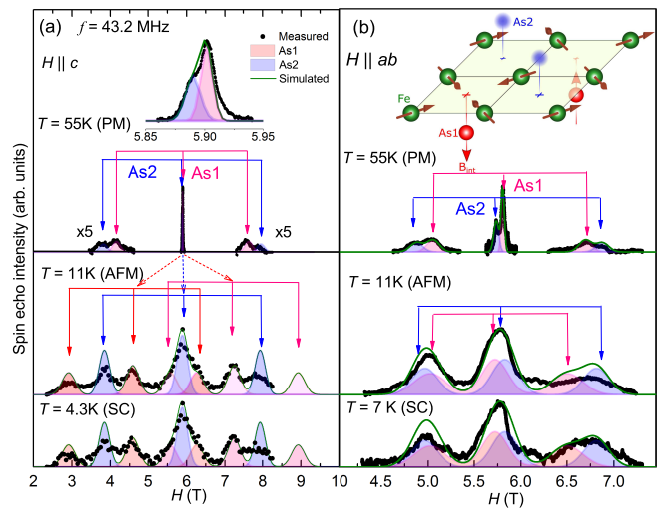} 
\caption{$T$ dependence of the field-swept $^{75}$As-NMR spectra of 4.9$\%$Ni-CaK1144 measured at $f$ = 43.2 MHz, for $H$ $\parallel$ $c$ axis (a) and $H$ $\parallel$ $ab$ plane (b). 
    The inset in (a) enlarges the central transition around 5.9 T. 
    The inset in (b) shows the sketch of hedgehog SVC spin structure on an Fe-As layer.
     The wine arrows represent the magnetic moments at the Fe sites and the red arrows represent the magnetic induction $B_{\rm int}$ at the As1 sites. 
 }
\label{fig:Spectra}
\end{figure}    

   Figures\ \ref{fig:Spectra} (a) and (b) show the $T$ dependence of field-swept $^{75}$As-NMR spectra of 4.9$\%$Ni-CaK1144 at $f$ = 43.2 MHz for two magnetic field directions, $H$ $\parallel$ $c$ axis and $H$ $\parallel$ $ab$ plane, respectively. 
   The typical spectrum for a nucleus with spin $I=3/2$ with Zeeman and quadrupolar interactions can be described by a nuclear spin Hamiltonian ${\cal{H}}=-\gamma\hbar(1+K)HI_z+\tfrac{h\nu_Q}{6}(3I_z^2-I^2)$,
where $H$ is the external field, $h$ is Planck's constant, $\hbar = h/2\pi$, $K$ is the Knight shift, and $\nu_{\rm Q}$ is the nuclear quadrupole frequency.
   The nuclear quadrupole frequency for an $I=3/2$ nuclei is given by $\nu_{\rm Q} = e^2QV_{\rm ZZ}/2h$, where $Q$ is the nuclear quadrupole moment and $V_{\rm ZZ}$ is the electric field gradient at the As site.
  When the Zeeman interaction is greater than the quadrupolar interaction, this Hamiltonian produces a spectrum with a sharp central transition line flanked by one satellite peak on either side. 
   The two inequivalent As sites of this 1144-structure results in two sets of $I=3/2$ quadrupole split lines which are actually observed for both $H$ directions in the paramagnetic state as seen in Fig. \ref{fig:Spectra}.
   The observed spectra are very similar to those in the pure CaK1144 where the lower field central peak with a greater Knight shift $K$ (and also larger $\nu_{\rm Q}$)  has been assigned to the As2 site and the higher field central peak with a smaller $K$ (and also smaller $\nu_{\rm Q}$) has been attributed to the As1 site \cite{Jean2017}.

\begin{figure}[tb]
\includegraphics[width=9.2cm]{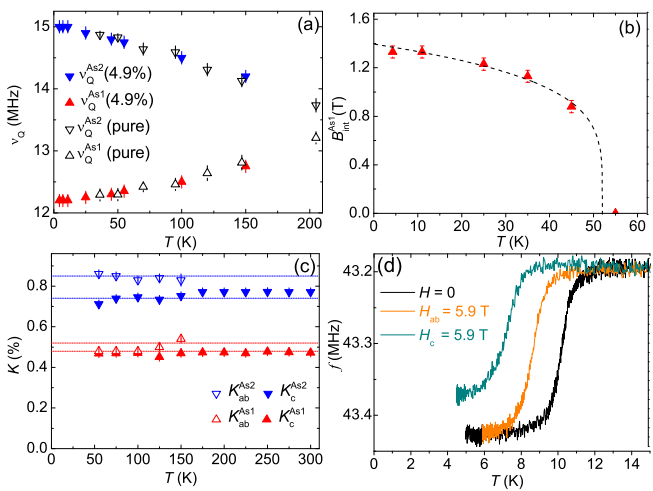} 
\caption{(a) $T$ dependences of quadrupole frequency $\nu_{\rm Q}$ for the As1 and the As2 sites, estimated from the NMR spectra. $\nu_{\rm Q}$s for pure CaK1144 are from Ref. \onlinecite{Jean2017}. (b) $T$ dependence of the internal magnetic induction $B_{\rm int}^{\rm As1}$ for the As1 site in the magnetic ordered state for $H$ $\parallel$ $c$ axis. The curve is a guide to the eyes. (c) $T$ dependences of the $^{75}$As-NMR shifts $K_{\rm c}$ and $K_{\rm ab}$ for the As1 and As2 sites. (d) $T$ dependence of the resonance frequency $f$ of the NMR tank circuit.
}
\label{fig:K}
\end{figure}

   Figure\ \ref{fig:K} shows the $T$ dependence of $\nu_{\rm Q}$, $K_{ab}$ ($H$ $\parallel$ $ab$) and $K_c$ ($H$ $\parallel$ $c$ axis) for the two As sites. 
   Due to the poor signal intensity at high $T$, $\nu_{\rm Q}$ and $K_{ab}$ can only be determined precisely up to 150 K. 
   For both the As sites, $K$s are nearly independent of $T$. 
   A similar weak $T$ dependence of $K$ has been observed in the non-magnetic CaK1144  \cite{Jean2017}, indicating that static magnetic susceptibility is almost insensitive to the electron doping and also to the magnetic or non-magnetic ground states.

   The $T$ dependences of $\nu_{\rm Q}$ of As1 and As2 of 4.9$\%$Ni-CaK1144 are similar to those of $\nu_{\rm Q}$ of the As sites in pure CaK1144 \cite{Jean2017} as shown in Fig.\ \ref{fig:K}(a).
     For the As1 site, with increasing $T$, $\nu_{\rm Q}$ increases from 12.2 MHz at 4.3 K to 12.75 MHz at 150 K, while the As2 site shows an opposite trend where $\nu_{\rm Q}$ decreases from 15.0 MHz at 4.3 K to 14.2 MHz at 150 K.
   The first-principles analysis shows the different $T$ dependences of $\nu_{\rm Q}$s for the two As sites can be explained by hedgehog SVC magnetic fluctuations \cite{Jean2017}.
  The full-width at half-maximum (FWHM) of the central line for $H$ $\parallel$ $c$ axis  is nearly independent of $T$ with $\sim$ 114 and $\sim$ 142 Oe for the As1 and As2 sites, respectively, in the paramagnetic state.
  The FWHMs of each satellite line are estimated to be 1.60 kOe and 1.65 kOe for the As1 and As2 sites at 55 K and  $H$ $\parallel$ $c$ axis, respectively.
   The linewidths of both the central and satellite lines are much greater than those in the pure CaK1144, which could be due to the disorder in the FeAs layer introduced by the substitution of Ni for Fe.

     When $T$ is lowered below $T_{\rm N}$ = 52 K, for $H$ $\parallel$ $c$ axis, each line of 4.9$\%$Ni-CaK1144  starts to broaden and the observed spectra become more complex, as typically shown in the middle panel of Fig.  \ref{fig:Spectra} (a).  
  As reported in Ref. \onlinecite{Meier20172}, the observed spectra in the magnetic state are well explained by the superposition of NMR spectrum from two As sites: six peaks from the As1 site (doubling the number of resonance lines due to internal magnetic induction $B_{\rm int}$) and three peaks from the As2 site.
   Here it is noted that the different  $\nu_{\rm Q}$ value for each As site makes unambiguous peak assignments of the complicated spectrum in the AFM state possible. 
    When $H$ is applied parallel to the $ab$ plane, in contrast, no splitting for the As1 NMR line is observed as shown in  Fig.  \ref{fig:Spectra} (b).
  The effective magnetic induction ${\bf B}_{\rm eff}$ is given by the vector sum of  ${\bf B}_{\rm int}$ at nucleus site and ${\bf H}$,   i.e., $|$$\bf{B}_{\rm eff}$$|$ = $|$$\bf{B}_{\rm int}$ + $\bf{H}$$|$.
    Therefore, when ${\bf B}_{\rm int}$ is parallel or antiparallel to $\bf{H}$,  $B_{\rm eff}$ = $H \pm B_{\rm int}$ and a splitting of each line is expected. 
    On the other hand, when $\bf{H}$ $\perp$ ${\bf B}_{\rm int}$, no splitting of the line is expected since  $B_{\rm eff}$ is expressed by $B_{\rm eff}$  =  $\sqrt {H^2+B_{\rm int}^2}$.
    Thus, the doubling the resonance lines at the As1 site only for $H$ $\parallel$ $c$ clearly shows that $B_{\rm int}$  at the As1 site is oriented along the $c$ axis. The absence of a clear splitting or shift of the resonance lines associated with the As2 site below $T_{\rm N}$ indicates the net $B_{\rm int}$ at the As2 sites is zero for both $H$ directions. This hyperfine field pattern is consistent with the hedgehog SVC [inset of Fig.\ \ref{fig:Spectra} (b)] \cite {O'Halloran2017,Meier20172}.
The line broadening indicates that $B_{\rm int}$ is slightly distributed probably originating from the distributions of the Fe ordered moments.
     It is  noted that no NMR signal from paramagnetic impurity phase can be observed, indicating a high quality of the sample.   

   Below $T_{\rm c}$, the NMR spectra measured for both $H$ directions in the SC state do not show any drastic change in shape.
   However, we observe a strong reduction of NMR signal intensity due to Meissner effect below $T_{\rm c}$.
   Here  $T_{\rm c}$s for both $H$ directions were determined by $\chi_{\rm ac}$ measurements.  
   As shown in Fig.\ \ref{fig:K}(d), $T_{\rm c}$  decreases to $\sim$  9 K and $\sim$  8 K for the application of $H$ = 5.9 T along the $ab$ plane and the $c$ axis, respectively, from  $T_{\rm c}$ $\sim$  10.5 K under $H$ = 0.
   The difference  in the reduction of $T_{\rm c}$ for the two $H$ directions is due to an anisotropy of upper critical field $H_{\rm c2}$, and the anisotropy parameter  $\gamma (T)$ = $H_{c2}^{ab}(T)$/$H_{c2}^c(T)$ is estimated to be $\sim$ 2 near $T_{\rm c}$, which is comparable to that in the pure CaK1144 \cite {Meier2016}.
   Since we do not observe any trace of paramagnetic impurity phases in the NMR spectrum below $T_{\rm N}$, we can exclude a possibility of phase separation in the compound. 
    Therefore we conclude the intrinsic coexistence of the hedgehog SVC and SC states.

    The $T$ dependence of $B_{\rm int}$ estimated from the splitting of the As1 central line for $H$ $\parallel$ $c$ is shown in Fig. \ref{fig:K}(b).  
    Just below $T_{\rm N}$, $B_{\rm int}$ starts to increase rapidly and saturates at low $T$, consistent with the second-order phase transition. 
    A similar $T$ dependence has been observed in the hyperfine field at the Fe sites from M\"ossbauer measurement \cite{Meier20172}.
    It should be noted that no significant reduction of $B_{\rm int}$ is observed below $T_{\rm c}$. 
    This indicates that SC for 4.9$\%$Ni-CaK1144 does not substantially disturb the hedgehog SVC AFM order parameter.
    For lower Ni substitution levels, when $T_{\rm c}$ is close to $T_{\rm N}$, this may change \cite{3Ni_data}.

\begin{figure}[tb]
\includegraphics[width=8.85cm]{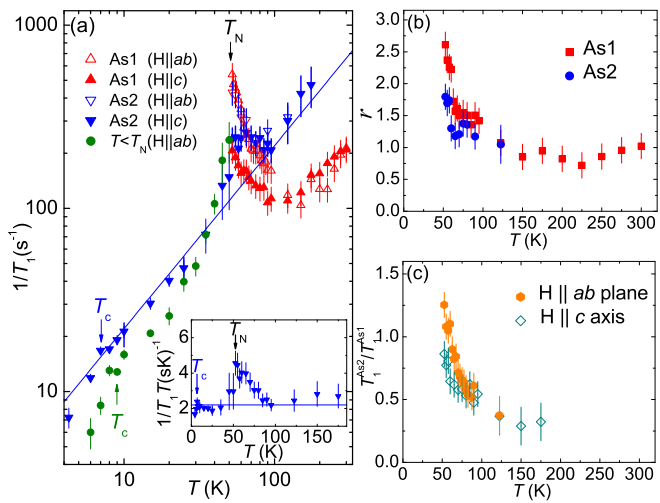} 
\caption{ (a) $T$ dependence of 1/$T_1$ of both As sites for $H$ $\parallel$ $c$ axis and $H$ $\parallel$ $ab$ plane. 
The inset shows the $T$ dependence of 1/$T_1T$ of As2 site for $H$ $\parallel$ $c$ axis.
The solid lines represent the Korringa relation 1/$T_1$ $\propto$ $T$. 
(b) $T$ dependence of $r \equiv T_{{\rm 1}c}/T_{{\rm 1}ab}$ for the As1 and A2 sites. 
(c) $T$ dependences of $T_{\rm 1}^{\rm As2}$/$T_{\rm 1}^{\rm As1}$ for $H$ $\parallel$ $c$ axis and $H$ $\parallel$ $ab$ plane.
 }
\label{fig:T1}
\end{figure}

   The coexistence of AFM and SC is further evidenced by 1/$T_{\rm 1}$ measurements.
   Figure \ \ref{fig:T1}(a) shows the $T$ dependences of 1/$T_{\rm 1}$ of As1 and As2 for $H$ $\parallel$ $c$ axis (1/$T_{1c}$; solid symbols)  and $H$ $\parallel$ $ab$ plane (1/$T_{1ab}$; open symbols).
   At high $T$ above $\sim$ 150 K, all 1/$T_1$s are nearly proportional to $T$, following the so-called Korringa relation.
    1/$T_1$ for As2 is greater than that of As1, which is due to  the different hyperfine coupling constant for each As site.
   With decreasing $T$, all $1/T_1$s start to increase below 100 K and then exhibit a peak at $T_{\rm N}$  = 52 K due to the hedgehog SVC order.
    Below $T_{\rm N}$, we were only able to measure 1/$T_1$ at As2 for $H \parallel c$ at the peak position where the NMR signal is dominated by the central-transition line of NMR spectrum for As2.  
     1/$T_1$ for As2 shows a sudden decrease just below $T_{\rm N}$ and then seems to follow the Korringa relation 1/$T_1$ $\propto$ $T$. 
      With a further decrease in $T$, 1/$T_1$ exhibits a decrease below $T_{\rm c}$ due to the SC transition, evidencing the coexistence of the AFM and SC states.
      The decrease in 1/$T_1$ below $T_{\rm c}$ is small, which is due to our limited temperature range of the measurements where NMR signal intensity becomes very poor at low temperatures below $T_{\rm c}$. 
      The strong reduction of the NMR signal intensity also evidences the observed NMR signals come from the SC region.
       It is interesting to point out that  the 1/$T_1T$ value below $T_{\rm N}$ is almost comparable to that of 1/$T_1T$ at higher $T$, as shown in the inset of  Fig. \ \ref{fig:T1}(a). 
      A similar $T$ dependence of $1/T_1T$ is observed in Ba(Fe$_{1-x}$Co$_x$)$_2$As$_2$ which exhibits the coexistence of the stripe-type AFM and SC states \cite{Cui2015}. 
      The value of 1/$T_1T$ at low $T$, comparable to that at high $T$,  indicates that significant AFM spin fluctuations persist in the AFM state \cite{Cui2015}.
       This is completely different from the case of Co-doped CaFe$_2$As$_2$, where the AFM spin fluctuations are strongly suppressed in the AFM state, and the AFM and SC do not coexistence \cite{Cui2015}.
       Therefore, we conclude that the remaining AFM spin fluctuations play an important role for the coexistence of AFM and SC in 4.9$\%$Ni-CaK1144.
      In particular, the AFM fluctuations in Ni-doped CaK1144 are characterized by the hedgehog-type spin correlations as shown below.
    
      We also attempted to measure 1/$T_1$ for $H \parallel ab$ in the AFM state for each As site separately. However,  it was very difficult to measure because of the overlap of the spectrum as shown in the Fig. \ref{fig:Spectra} (b). 
     Instead, we measured $1/T_1$ at the peak position to see how $T_1$ changes in the SC state.  
       Because of the overlap, the recovery behaviors of the nuclear magnetization were not reproduced  by the single  $T_{\rm 1}$ function, and a stretched  exponential function was used to extract 1/$T_1$ \cite{T1}.
        The olive circles show the $T$ dependence of 1/$T_1$, which shows a clear decrease below $T_{\rm c}$ without exhibiting a Hebel-Slichter coherence peak, providing again microscopic evidence of the coexistence of the hedgehog SVC AFM and SC states.
       No observation of the Hebel-Slichter coherence peak just below $T_{\rm c}$ is consistent with the data in the pure CaK1144 which exhibits an $s^\pm$ nodeless two-gap SC state \cite{Jean2017}.

    Finally we discuss the spin correlations in the paramagnetic state. 
  According to to previous NMR studies performed on Fe pnictides and related materials \cite{KitagawaSrFe2As2,SKitagawaAF-Fluctuation,FukazawaKBaFe2As2, Furukawa2014,Pandey2013,Ding2016}, the ratio $r$ $\equiv$ $T_{1c}$/$T_{1ab}$ depends on spin correlations. 
    In most of Fe pnictide SCs, $r$ is greater than unity corresponding to the AFM fluctuations with ${\bf Q}$ = ($\pi$,0) or  (0,$\pi$) \cite{KitagawaSrFe2As2,SKitagawaAF-Fluctuation,FukazawaKBaFe2As2, Furukawa2014}. 
   Here the wavevectors are given in the single-iron Brillouin zone notation. 
    On the other hand, $r$ = 0.5 is expected for ${\bf Q}$ = ($\pi$,$\pi$) spin correlations. 
    As plotted in Fig.\ \ref{fig:T1}(b), for both the As1 and As2 sites, $r$ is $\sim$ 1.0 at high $T$ and increases below $\sim$ 150 K down to $T_{\rm N}$. 
    This indicates that the AFM spin fluctuations characterized with  ${\bf Q}$ = ($\pi$,0) or (0,$\pi$) are enhanced below $\sim$ 150 K.
    It is noted that the $r$ ratio for the As2 site could be expected to be unity if the internal field at the As2 site is canceled out perfectly. 
The small enhancement of $r$ at the As2 site could be due to the distributions of either the Fe ordered moments or hyperfine coupling constant due to Ni substitutions. 
   In fact, such distributions will make the NMR spectrum broad, consistent with the observed spectra in the magnetic ordered state shown in Fig. 2.
     In most of  Fe-based SCs, the AFM spin fluctuations are considered to be the stripe-type spin correlations. 
     However, we show that the AFM spin fluctuations are characterized by the hedgehog-type AFM spin fluctuation in 4.9$\%$Ni-CaK1144.
     In the hedgehog SVC ordered state, the $B_{\rm int}$  at As1 is finite along the $c$ axis while the $B_{\rm int}$  at As2 is zero due to a cancellation originating from the characteristic spin structure \cite{Meier20172}.  
     Therefore, one expects that 1/$T_1$ for As1 is more enhanced than that for As2 if the AFM spin fluctuations originate from the hedgehog SVC-type spin correlations.  
      For stripe-type AFM fluctuations, the $T$ dependence of $1/T_1$ for As1 should scale to that of $1/T_1$ for As2 sites since there is no cancellation of the internal induction at either As site \cite{Meier20172}.  
     1/$T_{\rm 1}$ for As1 divided by 1/$T_{\rm 1}$ for As2 for $H$ $\parallel$ $ab$ and $H$ $\parallel$ $c$ are shown in Fig.\ \ref{fig:T1}(c).
     Above $\sim$ 150 K, the ratios of $T_1^{\rm As2}/T_1^{\rm As1}$ show a nearly $T$ independent value of $\sim$ 0.3 which could be determined by the different hyperfine coupling constants for the As1 and As2 sites. 
     Below $\sim$ 150 K,  clear enhancements of $T_1^{\rm As2}/T_1^{\rm As1}$ are observed. 
    This indicates that the As1 sites experience stronger AFM spin fluctuations than the As2 sites, evidencing the growth of the hedgehog SVC-type spin fluctuations.

     In summary, we have carried out  $^{75}$As NMR measurements to investigate the local electronic and magnetic properties of 4.9$\%$Ni-CaK1144.
     The new magnetic structure of the hedgehog spin vortex crystal with the $C_4$ symmetry has been clearly shown by  $^{75}$As NMR spectra observed below $T_{\rm N}$ $\sim$  52 K.
     The coexistence of the hedgehog SVC and SC states below $T_{\rm c}$ has been evidenced microscopically from the NMR spectrum and 1/$T_{\rm 1}$ measurements. 
     The hedgehog SVC-type spin correlations observed in the paramagnetic state are shown to persist in the AFM state, which is likely an important prerequisite for the  coexistence.    
     These results open up new avenues  for future research for understanding the relationship between the new antiferromagnetism and SC, and further detailed studies may provide some clues about the origin of high-$T_{\rm c}$ SC in iron-based SCs.

     We thank P. Wiecki, K. Rana, R. Fernandes, P. Orth, I. Mazin, A. Kreyssig,  V. Borisov and R. Valent\'i for helpful discussions. 
     The research was supported by the U.S. Department of Energy, Office of Basic Energy Sciences, Division of Materials Sciences and Engineering. Ames Laboratory is operated for the U.S. Department of Energy by Iowa State University under Contract No.~DE-AC02-07CH11358.
    WRM was supported by the Gordon and Betty Moore Foundations EPiQS Initiative through Grant GBMF4411.


\begin{thebibliography}{10}
\bibitem{Johnston2010} D. C. Johnston, Adv.  Phys. {\bf 59}, 803 (2010).
\bibitem{Canfield2010} P. C. Canfield and S.  L. Bud'ko, Annu. Rev. Condens. Matter Phys. {\bf 1}, 27 (2010).
\bibitem{Stewart2011} G.  R. Stewart, Rev. Mod. Phys. {\bf 83}, 1589 (2011).
\bibitem{Urbano2010} R. R. Urbano, E. L. Green, W. G. Moulton, A. P. Reyes, P. L. Kuhns, E. M. Bittar, C. Adriano, T. M. Garitezi, L. Bufaical, and P. G. Pagliuso, Phys. Rev. Lett. {\bf 105}, 107001 (2010).
\bibitem{Avci2011}S. Avci, O. Chmaissem, E. A. Goremychkin, S. Rosenkranz, J.-P. Castellan, D. Y. Chung, I. S. Todorov, J. A. Schlueter, H. Claus, M. G. Kanatzidis, A. Daoud-Aladine, D. Khalyavin, and R. Osborn, Phys. Rev. B {\bf 83}, 172503 (2011).
\bibitem{Wiesenmayer2011} E. Wiesenmayer, H. Luetkens, G. Pascua, R. Khasanov, A. Amato, H. Potts, B. Banusch, H.-H. Klauss, and D. Johrendt, Phys. Rev. Lett. {\bf 107},  237001 (2011).
\bibitem{Li2012} Z. Li, R. Zhou, Y. Liu, D. L. Sun, J. Yang, C. T. Lin, and G.-q. Zheng,  Phys. Rev. B {\bf 86}, 180501(R) (2012).
\bibitem{Iye2012} T. Iye, Y. Nakai, S. Kitagawa, K. Ishida, S. Kasahara, T. Shibauchi, Y. Matsuda, and T. Terashima, J. Phys. Soc. Jpn. {\bf 81}, 033701 (2012).
\bibitem{Pratt2009} D. K. Pratt, W. Tian, A. Kreyssig, J. L. Zarestky, S. Nandi, N. Ni, S. L. Bud'ko, P. C. Canfield, A. I. Goldman, and R. J. McQueeney,  Phys. Rev. Lett. {\bf 103}, 087001 (2009).
\bibitem{Ni2010} N. Ni, A. Thaler, J. Q. Yan, A. Kracher, E. Colombier, S. L. Bud'ko, P. C. Canfield, and S. T. Hannahs,  Phys. Rev. B {\bf 82}, 024519 (2010).
\bibitem{Ma2012} L. Ma, G. F. Ji, J. Dai, X. R. Lu, M. J. Eom, J. S. Kim, B. Normand, and W. Yu,  Phys. Rev. Lett. {\bf 109}, 197002 (2012).
\bibitem{Laplace2009} Y. Laplace, J. Bobroff, F. Rullier-Albenque, D. Colson, and A. Forget,  Phys. Rev. B {\bf 80}, 140501(R) (2009).
\bibitem{Cui2015}J. Cui, B. Roy, M. A. Tanatar, S. Ran, S. L. Bud'ko, R. Prozorov, P. C. Canfield and Y. Furukawa, Phys. Rev. B {\bf 92}, 184504 (2015).
\bibitem{Hoyer2016} M. Hoyer, R. M. Fernandes, A. Levchenko, and  J. Schmalian, Phys. Rev. B {\bf 93}, 144414 (2016).
\bibitem{Fernandes2016}R. M. Fernandes, S. A. Kivelson, and E. Berg, Phys. Rev. B {\bf 93}, 014511 (2016).
\bibitem{O'Halloran2017}J. O'Halloran, D. F. Agterberg, M. X. Chen, and M. Weinert, Phys. Rev. B {\bf 95}, 075104 (2017). 
\bibitem{Allred2016} J. M. Allred, K. M. Taddei, D. E. Bugaris, M. J. Krogstad, S. H. Lapidus, D. Y. Chung, H. Claus, M. G. Kanatzidis, D. E. Brown, J. Kang, R. M. Fernandes, I. Eremin, O. Chmaissem, and R. Osborn, Nat. Phys. {\bf 12}, 493 (2016).
\bibitem{Kim2010} M. G. Kim, A. Kreyssig, A. Thaler, D. K. Pratt, W. Tian, J. L. Zarestky, M. A. Green, S. L. Bud'ko, P. C. Canfield, R. J. McQueeney, and A. I. Goldman,  Phys. Rev. B {\bf 82}, 220503(R) (2010).
\bibitem{Avci2014} S. Avci, O. Chmaissem, J. M. Allred, S. Rosenkranz, I. Eremin, A. V. Chubukov, D. E. Bugaris, D. Y. Chung, M. G. Kanatzidis, J.-P. Castellan, J. A. Schlueter, H. Claus, D. D. Khalyavin, P. Manuel, A. Daoud-Aladine, and R. Osborn, Nat. Commun. {\bf 5}, 3845 (2014).
\bibitem{Wang2016} L. Wang, F. Hardy, A. E. B\"ohmer, T. Wolf, P. Schweiss, and C. Meingast, Phys. Rev. B {\bf 93}, 014514 (2016).
\bibitem{Hassinger2012} E. Hassinger, G. Gredat, F. Valade, S. R. de Cotret, A. Juneau-Fecteau, J.-Ph. Reid, H. Kim, M. A. Tanatar, R. Prozorov, B. Shen, H.-H. Wen, N. Doiron-Leyraud, and L. Taillefer,  Phys. Rev. B {\bf 86}, 140502(R) (2012).
\bibitem{Bohmer2015} A. E. B\"ohmer, F. Hardy, L. Wang, T. Wolf, P. Schweiss, and C. Meingast, Nat. Commun. {\bf 6}, 7911 (2015).
\bibitem{Allred2015} J. M. Allred, S. Avci, D. Y. Chung, H. Claus, D. D. Khalyavin, P. Manuel, K. M. Taddei, M. G. Kanatzidis, S. Rosenkranz, R. Osborn, and O. Chmaissem,  Phys. Rev. B {\bf 92}, 094515 (2015).
\bibitem{Hassinger2016} E. Hassinger, G. Gredat, F. Valade, S. R. de Cotret, O. Cyr-Choini\`ere, A. Juneau-Fecteau, J.-Ph. Reid, H. Kim, M. A. Tanatar, R. Prozorov, B. Shen, H.-H. Wen, N. Doiron-Leyraud, and L. Taillefer, Phys. Rev. B {\bf 93}, 144401 (2016).
\bibitem{Meier20172} W. R. Meier, Q.-P. Ding, A. Kreyssig, S. L. Bud'ko, A. Sapkota, K. Kothapalli, V. Borisov, R. Valent\'i, C. D. Batista, P. P. Orth, R. M. Fernandes, A. I. Goldman, Y. Furukawa, A. E. B\"ohmer, and P. C. Canfield, arXiv:1706.01067.
\bibitem{Iyo2016} A. Iyo, K. Kawashima, T. Kinjo, T. Nishio, S. Ishida, H. Fujihisa, Y. Gotoh, K. Kihou, H. Eisaki, and Y. Yoshida, J. Am. Chem. Soc. {\bf 138}, 3410 (2016).
\bibitem{Meier2016}  W. R. Meier, T. Kong, U. S. Kaluarachchi, V. Taufour, N. H. Jo, G. Drachuck, A. E. B\"ohmer, S. M. Saunders, A. Sapkota, A. Kreyssig, M. A. Tanatar, R. Prozorov, A. I. Goldman, Fedor F. Balakirev, Alex Gurevich, S. L. Bud'ko, and P. C. Canfield, Phys. Rev. B {\bf 94}, 064501 (2016).
\bibitem{ARPES2016} D. Mou, T. Kong, W. R. Meier, F. Lochner, L.-L. Wang, Q. Lin, Y. Wu, S.  L. Bud'ko, I. Eremin, D.  D. Johnson, P.  C. Canfield, and A. Kaminski, Phys. Rev. Lett. {\bf 117},  277001 (2016).
\bibitem{Lochner2017} F. Lochner, F. Ahn, T. Hickel, and I. Eremin, Phys. Rev. B  {\bf96}, 094521 (2017).
\bibitem{uSR2017} P. K. Biswas, A. Iyo, Y. Yoshida, H. Eisaki, K. Kawashima and A. D. Hillier, Phys. Rev. B  {\bf95}, 140505 (2017).
\bibitem{Cho2017} K. Cho, A. Fente, S. Teknowijoyo, M. A. Tanatar, K. R. Joshi, N. M. Nusran, T. Kong, W. R. Meier, U. Kaluarachchi, I. Guillam\'on, H. Suderow, S. L. Bud'ko, P. C. Canfield and R. Prozorov, Phys. Rev. B {\bf 95}, 100502 (2017).
\bibitem{Jean2017} J. Cui, Q.-P. Ding, W. R. Meier, A. E. B\"ohmer, T. Kong, V. Borisov, Y. Lee, S. L. Bud'ko, R. Valent\'i, P. C. Canfield, and Y. Furukawa, Phys. Rev. B  {\bf96}, 104512 (2017).
\bibitem{Iida2017} K. Iida, M. Ishikado, Y. Nagai, H. Yoshida, A. D. Christianson, N. Murai, K. Kawashima, Y. Yoshida, H. Eisaki, and A. Iyo, J. Phys. Soc. Jpn. {\bf 86}, 0937030 (2017).
\bibitem{Meier2017} W. R. Meier, T. Kong, S. L. Bud'ko, and P. C. Canfield, Phys. Rev. Materials {\bf 1}, 013401 (2017).
\bibitem{T1}  See Supplemental Material for representative nuclear magnetization recovery curves and details for the  $1/T_1$ fitting.
\bibitem{3Ni_data} 
The suppression of AFM order below $T_{\rm c}$  is due to the competition between AFM and SC orders. In 4.9$\%$Ni-CaKFe$_4$As$_4$, $T_{\rm c}$ is less than 20$\%$ of $T_{\rm N}$, the suppression of AFM order parameter may be very small, which may not be able to detect within our experimental uncertainty. 
When the ratio of  $T_{\rm c}$/$T_{\rm N}$ increases, a larger reduction of the AFM order parameter below $T_{\rm c}$ was expected.
    Our preliminary NMR spectrum measurements on 3.3$\%$Ni-doped CaK(Fe$_{0.967}$Ni$_{0.033}$)$_4$As$_4$  ($T_{\rm N}$ $\sim$ 45 K, $T_{\rm c}$ $\sim$ 23 K) show a small reduction of the internal magnetic induction at the As1 site below $T_{\rm c}$.  
\bibitem{KitagawaSrFe2As2} K. Kitagawa, N. Katayama, K. Ohgushi, and M. Takigawa, J. Phys. Soc. Jpn. {\bf 78}, 063706 (2009).
\bibitem{SKitagawaAF-Fluctuation} S. Kitagawa, Y. Nakai, T. Iye, K. Ishida, Y. Kamihara, M. Hirano, and H. Hosono, Phys. Rev. B {\bf 81}, 212502 (2010). 
\bibitem{FukazawaKBaFe2As2} M. Hirano, Y. Yamada, T. Saito, R. Nagashima, T. Konishi, T. Toriyama, Y. Ohta, H. Fukazawa, Y. Kohori, Y. Furukawa, K. Kihou, C.-H. Lee, A. Iyo and H. Eisaki, J. Phys. Soc. Jpn. {\bf 81}, 054704  (2012). 
\bibitem{Furukawa2014} Y. Furukawa, B. Roy, S. Ran, S. L. Bud'ko, and P. C. Canfield, Phys. Rev. B {\bf 89}, 121109(R) (2014).
\bibitem{Pandey2013} A. Pandey, D. G. Quirinale, W. Jayasekara, A. Sapkota, M. G. Kim, R. S. Dhaka, Y. Lee, T. W. Heitmann, P. W. Stephens, V. Ogloblichev, A. Kreyssig, R. J. McQueeney, A. I. Goldman, A. Kaminski, B. N. Harmon, Y. Furukawa, and D. C. Johnston, Phys. Rev. B {\bf 88}, 014526(2013).
\bibitem{Ding2016} Q.-P. Ding, P. Wiecki, V.  K. Anand, N.  S. Sangeetha, Y. Lee, D.  C. Johnston, and Y. Furukawa, Phys. Rev. B {\bf 93}, 140502 (2016).


\end{thebibliography}
\end{document}